\documentclass[proceedings]{rmaa}

\usepackage{rmaacite}

\renewcommand{\P}[1]{%
\ifnum#1=1\hbox{OW~168--326E}\fi
\ifnum#1=2\hbox{OW~167--317}\fi
\ifnum#1=3\hbox{OW~163--317}\fi
\ifnum#1=5\hbox{OW~158--323}\fi
\ifnum#1=0\hbox{OW~171--334}\fi}

\title{Hidden Blazars and Emission Line Variability of High Redshift Quasars}
\author{Feng Ma
  \affil{Astronomy Department, the University of Texas at Austin} }

\fulladdresses{
\item Feng Ma: Astronomy Department, the University of Texas at Austin, 
Austin, TX 78712-1083, USA (feng@astro.as.utexas.edu).}

\shortauthor{Ma}
\shorttitle{Hidden Blazars and Quasars}


\abstract{ We have carried out a survey to search for hidden blazars
in a sample of $z{\sim}2$ radio-loud quasars. The idea is based
on our prediction that we should be able to see large 
CIV line variability not associated with observed continuum
variations or most other emission lines in every radio-loud
quasar. Here we report the initial results including the 
discovery of large CIV line variations
in two quasars.}

\listofauthors{F.~Ma}
\indexauthor{Ma, F.}

\begin{document}

\maketitle

\section{Introduction}
\label{sec:intro}

Jet-disk systems may exist in many astronomical objects such as
gamma-ray bursters, proto-stars and radio-loud quasars. 
Beamed emission associated with jets is often not observable
if the beam is pointing away from the observer. 
A belief in a 
simple representation  of nature
leads people to seek unification, adopting, for
radio-loud quasars, 
a ``unified scheme''. 
With increasing viewing angle from the jet, 
we see blazars, core--dominant quasars, lobe--dominant quasars or
radio galaxies.  If the unified scheme is correct, 
every quasar should harbor a blazar even though we do not see
most of them because they are beamed away from the line of sight. 
Blazars outburst every $\sim$10 years (Fan et al. 1999), and 
each burst last $\sim$1 year. If we assume the blazar 
beam illuminates 2\% of the gas in Broad Emission Line Region (BELR),
we were able to predict (Ma \& Wills 1998): 
1. collisionally excited lines such as SiIV $\lambda$1397 and 
CIV $\lambda$1549 are strongly enhanced by an outbursting blazar (Fig. 1); 
2. other emission lines including Ly$\alpha$ and CIII] $\lambda$1909
are little affected; 
3. for any radio--loud quasars we should be able to see
CIV and SiIV lines increase by over 50\% once every $\sim$30 years
(for $z{\sim}2$), and they last $\sim$3 years; 
4. if we examine a sample of radio--loud quasars, $\sim$10\% of 
them should show stronger CIV and SiIV, i.e., their hidden blazars are
in outburst. This phenomenon is distinguishable from 
emission line variations responding to ``normal'' (non--blazar) AGN
continuum in that  Ly$\alpha$, CIV$\lambda$1549 and CIII] $\lambda$1909
vary at similar amplitude responding to the ionizing continuum 
(Kaspi \& Netzer 1999).

\begin{figure}[ht]
  \begin{center}
    \leavevmode
    \includegraphics[width=\textwidth,height=11cm]{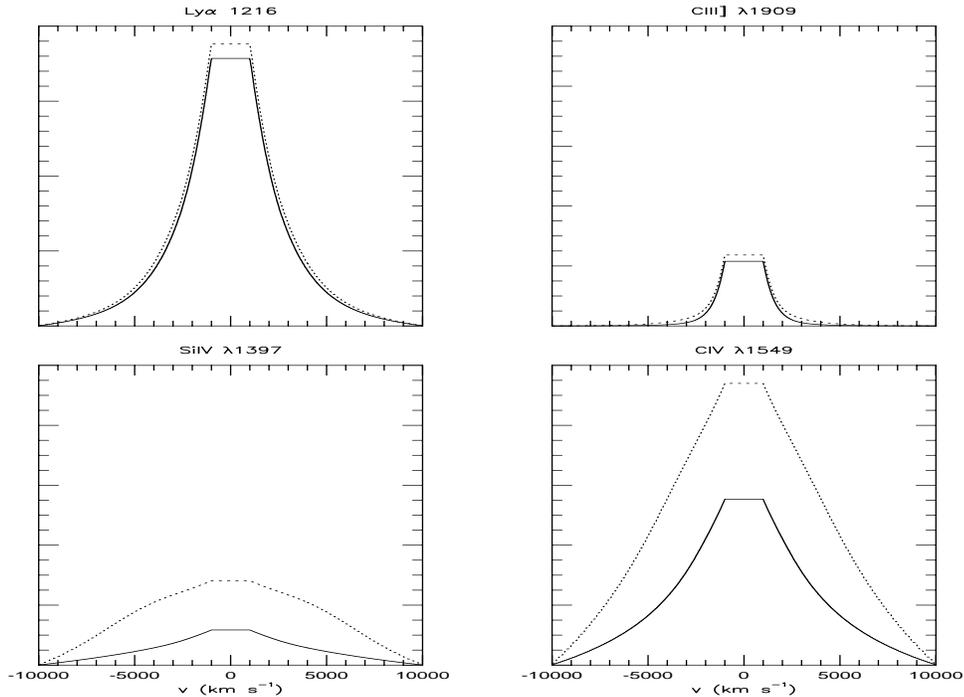}
\caption{Theoretical prediction of emission line variations
during an outburst of the hidden blazar. Ly$\alpha$ and CIII]
$\lambda$1909 are little affected by the hidden blazar, while
SiIV $\lambda$1397 and CIV $\lambda$1549 are dramatically 
enhanced. }
    \label{fig:dissprofp.ps}
  \end{center}
\end{figure}

\section{Observations}

Using the Large Cassegrain 
Spectrograph on the 2.7--m Harlan J. 
Smith telescope at McDonald Observatory, 
we have observed the spectra of 53 radio-loud quasars
of $z{\sim}2$, and have compared  45 of them with
historical data taken over 10 years ago. 
The 45 objects in our sample can be classified as five classes (Table 1). 
The flux is not always absolutely calibrated, and the difference 
in continuum levels is likely to be the result of clouds or 
wavelength-dependent
slit losses caused by atmospheric dispersion. 
We scale two spectra to match their continua, and the
line ratio is preserved during the scaling unless there
is an (unreasonably) large variation in the continuum shape. 
Some spectra from our sample are given in Fig. 2.  
PKS 0038$-$019 ($z=1.67$) and MRC 0238+100 ($z=1.83$) 
are both lobe--dominant quasars and show little optical 
variability from Digitized Sky Surveys I \& II, and hence
their CIV line variations are unlikely caused by large 
continuum shape change. 

\begin{table}
\caption{Classification of objects in the sample. }
\begin{center}
\begin{tabular}{crrr}\hline\hline
Class & \# & Description & Interpretation\\
\hline
A  &  2  & $>$30\% variations in CIV  & outbursting hidden blazars uncovered   \\
\\
B  &  7  & $\sim$15\% variations in CIV &  hidden blazars or CIV more variable  \\
\\
C  &  20 & no variations in any lines & high-$z$ and high-$L$  quasars less variable; \\
   &     &                    & observations and comparison are valid\\
\\
D  &  14 & all lines ``vary'' in proportion  & continuum difference at different epochs; \\
   &     &                   & line response to normal AGN continuum\\
\\
E  &  2  & Ly$\alpha$ show $\sim$15\% variations  & continuum more variable in the blue\\
\hline
\end{tabular}
\end{center}
\end{table}

\begin{figure}[tp]
  \begin{center}
    \leavevmode
    \includegraphics[width=\textwidth,height=19cm]{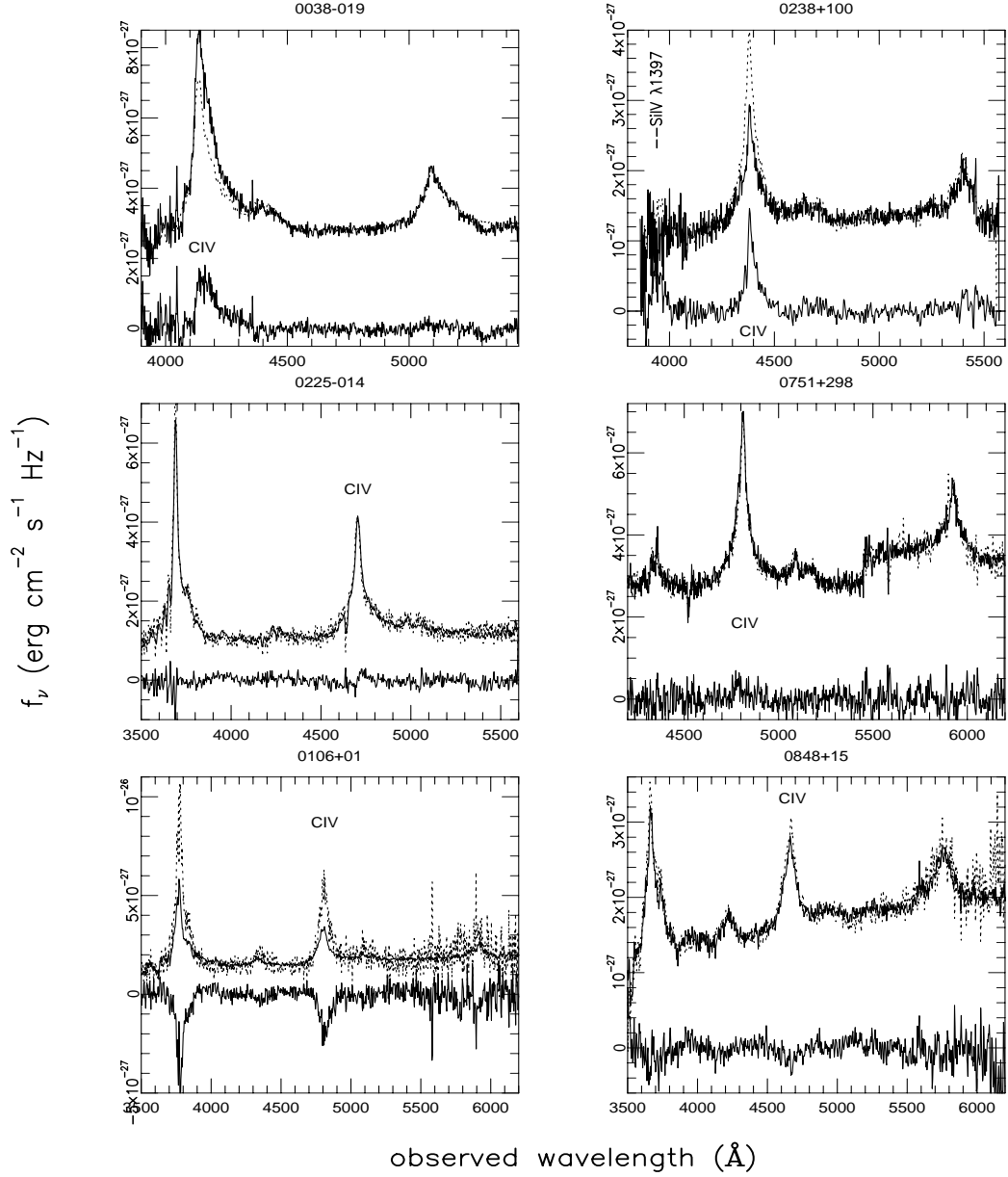}
\caption{The continua at different epochs have
been scaled to the same
level. The differences of the spectra are
also shown. 0038$-$019 and 0238$+$100 show large CIV variations (class A) 
when comparing spectra in  1999 (dotted lines) and in 1986 (solid lines, 
Barthel et al. 1990).
0225$-$014 and 0751$+$298 are examples of class C whose
spectra taken over a time interval of 10 years are in excellent agreement.
The apparent continuum jump in 0751$+$298 is due to the
different flux levels in the red and blue channel of
a double spectrograph. The Ly$\alpha$/CIV ratios in
0106$+$01 and 0848$+$15 at the two epochs remain unchanged and they
belong to class D.}
    \label{fig:mcd2000}
  \end{center}
\end{figure}

\section{Discussions}

Our discovery of two hidden blazars via emission line variability 
strongly supports the unified scheme.  
The idea in this work suggests a new way to look 
for hidden beamed emission via reprocessed, more
isotropic line emission, which can be applied to other jet--disk systems
such as gamma-ray bursters. The observations suggest that 
there is BELR gas in the polar regions 
and thus challenges the disk--wind model (Murray et al. 1995) 
for the BELR. While 
the stellar model (Alexander \& Netzer 1994) has been challenged by 
line profile studies (Arav et al. 1998),  we may need a new 
source responsible for at least a bulk of gas in BELR. 
We consider a BELR made of tidally disrupted stars
(Shields 1989, Roos 1992) with winds blowing off the
tidal streams by the radiation pressure. This model gives
satisfactory line ratios and line profiles (Ma 2000). 

In this model the tidal forces stretch the disrupted stars with a large
and continuous distribution in velocity and density,
solving the discrete clouds model problems such as confinement.
A rotating black hole will eject stars very differently, due to the flattened
event horizon and local dragging of inertia frames. Consequently, this may
explain the differences between radio--loud and radio--quiet quasars.
Most of the BELR parameters are determined by the central black hole
mass and the disruption process, and hence there is no need to
introduce free parameters such as density distribution and covering
factor. The winds from the tidal streams have comparable
covering factor compared with the tidal streams themselves,
hence lowering the requirement of high tidal disruption rate.
The winds and the tidal streams have very
distinct density ($10^{6-8}$cm$^{-3}$ vs. $10^{10-13}$cm$^{-3}$
at {\it similar} spatial location. The winds also have higher
velocities. The two components give different line ratios and 
profiles for different lines, and may solve long standing 
puzzles such as MgII deficiency. 
Finally, this model helps drawing a 
picture of quasar-galaxy connection, 
that AGN activity may be triggered in only 1\% of all galaxies 
during galaxy mergers. The merging boosts the tidal 
disruption rate dramatically from $10^{-4}$yr$^{-1}$ to
$\sim$ 1 yr$^{-1}$ (see, e.g., Roos 1992, and references therein). 
The bound part of tidal remnants offers a bulk 
of material to the accretion disk while the ejected unbound part 
supplies the BELR. 
We note that there could be more than one source  for BELR gas. 
Besides the tidal remnants, accretion disk winds, stellar 
atmosphere and stellar winds, remnants from stellar collisions
and star--disk interactions may all play a role in emitting
broad lines even though none of them is a solely responsible for
BELR. 

\acknowledgments

I thank Bev Wills, Derek Wills, and Greg Shields for their input to this
work. Jack Baldwin, Peter Barthel, Paul Francis, and Marianne Vestergaard
made their data available in digital form. We are also grateful to
David Doss for help
with the observations and Gary Ferland for making his code
CLOUDY available.

\end{document}